# Localized spin-orbit polaron in the magnetic Weyl semimetal Co$_3$Sn$_2$S$_2$

*Yuqing Xing[1,2,3†], Jianlei Shen[1,2†], Hui Chen[1,2,3†], Li Huang[1,2,3†], Yuxiang Gao[1,2,3], Qi Zheng[1,2,3], Yu-Yang Zhang[2,3], Geng Li[1,2,3], Bin Hu[1,2,3], Guojian Qian[1,2,3], Lu Cao[1,2,3], Xianli Zhang[1,2,3], Peng Fan[1,2,3], Ruisong Ma[1,2,3], Qi Wang[4], Qiangwei Yin[4], Hechang Lei[4], Wei Ji[4], Shixuan Du[1,2,3,5], Haitao Yang[1,2,3], Chengmin Shen[1,2,3], Xiao Lin[2,1,3], Enke Liu[1,2,5\*], Baogen Shen[1,2], Ziqiang Wang[6\*], and Hong-Jun Gao[1,2,3,5\**]*

[1] *Beijing National Center for Condensed Matter Physics and Institute of Physics, Chinese Academy of Sciences, Beijing 100190, PR China*

[2] *School of Physical Sciences, University of Chinese Academy of Sciences, Beijing 100190, PR China*

[3] *CAS Center for Excellence in Topological Quantum Computation, University of Chinese Academy of Sciences, Beijing 100190, PR China*

[4] *Beijing Key Laboratory of Optoelectronic Functional Materials & Micro-Nano Devices, Department of Physics, Renmin University of China, Beijing 100872, PR China*

[5] *Songshan Lake Materials Laboratory, Dongguan, Guangdong 523808, PR China*

[6] *Department of Physics, Boston College, Chestnut Hill, MA, USA*

†These authors contributed equally to this work

\*Correspondence to: hjgao@iphy.ac.cn, wangzi@bc.edu, ekliu@iphy.ac.cn

Abstract

The kagome lattice $Co_3Sn_2S_2$ exhibits the quintessential topological phenomena of a magnetic Weyl semimetal such as the chiral anomaly and Fermi-arc surface states. Probing its magnetic properties is crucial for understanding this correlated topological state. Here, using spin-polarized scanning tunneling microscopy/spectroscopy (STM/S), we report the discovery of localized spin-orbit polarons (SOPs) with three-fold rotation symmetry nucleated around single-S vacancies in $Co_3Sn_2S_2$. The SOPs carry a spin-polarized magnetic moment and a large orbital magnetization of a topological origin associated with the Berry phase and the persistent circulating current. Appreciable magneto-elastic coupling of the SOP is detected by atomic force microscope and STM. Our findings suggest that the SOPs can enhance magnetism and stability of the magnetic Weyl nodes for more robust time-reversal-symmetry-breaking topological phenomena. Controlled engineering of the SOPs may pave the way toward practical applications in functional quantum devices.

As a magnetic Weyl semimetal [1–4], $Co_3Sn_2S_2$ integrates the flat bands of correlated *d*-electrons on the kagome lattice [5–7] into the topological electronic structure [8–11] and exhibits remarkably novel phenomena such as the giant anomalous Hall/Nernst effect [8,12], chiral-anomaly [8], surface-termination dependent Fermi arcs [11], topological catalysis [13], flat band Berry phase and orbital magnetization [14]. While playing a fundamental role in driving these intriguing properties by breaking the time-reversal symmetry, the nature of the magnetism and the associated physics of the correlated *d*-electrons in the kagome lattice $Co_3Sn_2S_2$ are not understood. Probing the nature of the magnetism in $Co_3Sn_2S_2$ is thus crucial for understanding the physical origin, and for the control of the emergent quantum states via manipulations of magnetic order [15–17].

Here, we study localized magnetic polarons nucleated around single atomic S vacancies on S-terminated surface in $Co_3Sn_2S_2$ by spin-polarized STM. They emerge as bound states in the conductance map with a three-fold rotation symmetry. Applying external magnetic fields up to ±6 T normal to the surface reveals that the binding energy of the localized magnetic polaron linearly increases as a function of the field magnitude regardless of the field direction. This anomalous Zeeman response of a magnetic bound state has not been observed before and indicates dominant orbital magnetization contribution to the local magnetic moment (~ 1.35 $\mu_B$). Appreciable magneto-elastic coupling is also detected near the S-vacancy. We term this new excitation as a localized spin-orbit polaron (SOP) and argue that the large orbital magnetization has a topological origin associated with the Berry phase and the persistent circulating current due to the magnetoelectric effect of the topological magnet.

$Co_3Sn_2S_2$ has a layered structure which consists of two hexagonal planes of S and Sn as well as a $Co_3Sn$ kagome layer sandwiched between the S atoms (Fig. 1a-b). It is a ferromagnet with a Curie temperature of 177 K and a low-temperature out-of-plane magnetic moment of about 0.3 $\mu_B$/Co [9,18,19]. Single crystals of $Co_3Sn_2S_2$ were cleaved in-situ at 6 K under ultrahigh-vacuum and immediately transferred to the scanning tunneling microscope (STM) chamber. Weaker bonds between S and Sn layers offer a cleave plane and possibly lead to S-terminated and Sn-terminated surfaces (*11*). In STM topographic images, two types, i.e. Type-I and Type-II, of cleaved surfaces were observed. The Type-I surface shows a hexagonal-like lattice with randomly distributed vacancies (Fig. S1a), while the Type-II surface exhibits a similar lattice with adatoms and clusters (Fig. S1c). Combined with work function measurements using a noncontact atomic force microscope (AFM) and DFT calculations, we determined that the surface with the higher work function (the Type-I surface) is the S-terminated surface (see Supplementary Text I, Fig. S1 and Fig. S2).

We next studied the properties of localized excitations by focusing on a region with S vacancies on the S-terminated surface. A large-scale topographic image shows randomly disturbed S vacancies of the focused region (Fig. 1c) and a zoomed-in image depicts a single-S vacancy (Fig. 1e). A typical off-vacancy dI/dV spectrum (black curve in Fig. 1d) closely resembles the one taken in a region free of S vacancies on the S-surface, exhibiting an energy range of suppressed and flat density of states of about 300 meV [11,20], and a broad hump around +50 meV, which originates from the topological surface states of the $Co_3Sn_2S_2$ magnetic Weyl semimetal [11,20]. There is also a peak sitting at the edge of the valence band at -350 meV. Differently, the dI/dV spectrum taken at the S-vacancy (orange curve in Fig. 1d) shows suppressed density of states around +50 meV and -350 meV. Meanwhile, a series of approximately equal-spaced spectral peaks emerge just above the valence band inside the region of suppressed density of states, indicating bound states formation at the S-vacancy. From statistical analysis, we determine the average energy spacing between the spectral peaks to be ~16 meV (Fig. S3).

Spatial distributions of these spectroscopic features were recorded in dI/dV maps (Fig. 1f-j and Fig. S4). The conductance map at +50 mV (Fig. 1f) shows an atomically modulated scattering pattern of the low-energy surface states at the S vacancy. The map acquired at -350 mV (Fig. 1g) reveals a three-fold structure with suppressed intensity around the S-vacancy and all its six neighboring S atoms. Remarkably, the conductance maps (Fig. 1h-j) recorded at those three discernable dI/dV peaks, i.e. at -322, -300, and -283 mV, all show localized flower-petal shaped patterns. Each pattern exhibits a three-fold rotation symmetry centered at the S-vacancy and the boundary of the pattern reaches as far as the six neighboring S atoms, which approximately defines the size of the bound states.

The peak residing at -283 mV (Fig. 1d) is the sharpest and the most localized one, which is referred to the primary bound state. It consists of two sub-peaks with small energy splitting (Fig. S5). The brightest portion of its intensity is distributed over the three "triangles" in each of the three petals that corresponds to the three up-triangles in the underlying Co kagome lattice (Fig. 1j), clearly demonstrating the driving force for the formation of the bound state polaron is the localization of the Co $d$-electrons by the vacancy potential of the missing $S^{2-}$ ion, which hybridize with the S $p$-electrons. Interestingly, the conductance map of the -283 mV state (Fig. 1j), also those of the -322 and -300 mV states (Fig. 1h and 1i), show a dramatic contrast/intensity reversal in comparison with the map of the -350 mV state (Fig. 1g), suggesting the -350 mV state be an antibound state, an affirmation of the bound state nature of the S-vacancy induced polaron. A ~67 meV energy separation between the bound and antibound states allows us to estimate the "bandwidth" of the polaronic state.

To investigate the magnetic properties of the bound states, we used spin-polarized Ni tips to measure dI/dV spectra on the single-S vacancy (Fig. 2a). The spin-polarization of the Ni tip was calibrated on Co/Cu(111) (Fig. S6). An external magnetic field of +0.6 T (-0.6 T) was applied to magnetize the Ni tip to be in the spin up (spin down) state. This tip was then used to measure the spin-dependent dI/dV spectra at 0 T (Supplementary Text II). We did not observe appreciable spin-polarized contrast in any defect-free regions of the S-terminated surface, indicating that the intact S-terminated surface is nearly non-magnetic (Fig. 2b). At the S vacancy, the dI/dV spectra, however, show strong magnetic contrast from approximately -350 mV to -280 mV where the bound states reside, indicating that the bound states are magnetic with a spin-down majority (Fig. 2c).

Figure 2d shows the spin flip operation of the Ni tip [21,22], where we zoomed in to more closely show the primary bound state at ~-283 mV (marked by the black arrow in Fig. 2c). Particularly, a spin-down tip was initially prepared, which gave a pronounced peak around -283 mV (left panel in Fig. 2d). The polarization of the tip was then flipped to spin-up by an external magnetic field of +0.6 T. Given this tip, the intensity of the -283 mV peak reduces while the peak position keeps unchanged (middle panel in Fig. 2d). After flipping the tip spin back to spin-down, the peak intensity restores (right panel in Fig. 2d). This result demonstrates that the bound states are magnetic polarons introduced by the S vacancies (Fig. S7a). Indeed, from the high resolution dI/dV map acquired with the W tip around the S vacancy (Fig. 2e), the three-fold symmetric spatial profile of the bound magnetic polaron can be traced out and superimposed onto the atomic structure projected to the S-surface (Fig. 2f), revealing its overall correlation with the underlying Co atoms.

We further investigated the nature of the localized magnetic polaron by measuring the magnetic field response of the spectral peaks using a normal W tip. The field is applied perpendicularly to the sample surface, ranging from -6 T to 6 T. The energies of the both sub-peaks of the primary bound state shift linearly toward the higher energy side ($\Delta E > 0$), *independent of the direction of the magnetic field.* This unusual behavior is reproducible on various individual S vacancies (Fig. S8). The Zeeman effect $\Delta E = -\vec{\mu} \cdot \vec{H}$ indicates that the energy of a polarized (nondegenerate) magnetic state would decrease if the magnetic moment ($\vec{\mu}$) was parallel to the applied field but would increase if it was oriented antiparallel. The observed anomalous Zeeman response has two important physical implications. A $\Delta E$ independence of field directions implies that the magnetic field couples to the magnetization of the bound state, which simultaneously flips upon reversing the field direction. In addition, $\Delta E > 0$ indicates that the net magnetization is always oriented antiparallel to the direction of the magnetic field, indicative of a dominant orbital (diamagnetic) contribution, including the pseudospin (electron spin + atomic orbital) and the "molecular" orbital magnetization over the cluster of atoms in the presence of spin-orbit coupling.

To emphasize its orbital magnetic moment, we dub the magnetic bound state as a localized spin-orbit polaron (SOP). Fitting the two sub-peak positions as a linear function of the magnetic field (Fig. 3c), we obtained a slope of 75 μeV/T = 1.35 $\mu_B$ for the effective magnetic moment of the SOP. The anomalous Zeeman response was recently observed for the flat portion of the itinerant band states corresponding to a peak in the density of states at low energies and attributed to the Berry phase induced orbital magnetization [14]. A crucial difference of the previous work from the present lies in that the energy shift confined to the flat part of the band states does not correspond to a net magnetic moment and the orbital magnetization averages out to approximately zero when all momentum states are considered [14]. The anomalous Zeeman shift observed here, however, comes from a localized bound state and corresponds to the overall magnetic field response of a net physical magnetic moment. The large orbital magnetization of the SOP most likely originates from the persistent circulating current around the S-vacancy due to the Berry phase and magnetoelectric effect of the topological Weyl semimetal (Supplementary Text III).

Finally, we considered the magneto-elastic coupling within the localized SOP. The lattice distortion around the S vacancies measured by a noncontact AFM (Fig. 3d) and STM (Fig. 3e) shows appreciable local atomic displacements at 0 T. The measured average nearest atomic distances around the S vacancy determine the local atomic displacement ratio as its percentage change from the average distance of an intact region. The displacement ratio significantly decreases with increasing strength of the field applied along the *c*-axis (Fig. S9). Remarkably, one-third of the displacement ratio can be manipulated by a magnetic field up to 6 T. These observations further support the SOP nature of the bound states and the strong magneto-elastic coupling in $Co_3Sn_2S_2$.

The discovery of the localized SOP (Fig. 3f) opens a novel route for manipulating the magnetic order and the topological phenomena in Weyl semimetal $Co_3Sn_2S_2$. The STM observed S vacancies on S-terminated surfaces reflect the presence of bulk S vacancies, where localized SOPs are expected to nucleate with similar physical properties. DFT calculations reveal that the bulk vacancies are magnetic and significantly enhance the magnetic moment of neighboring Co atoms (Fig. S7b). The density of the SOP can thus be controlled by that of the S deficiency by varying S pressure and temperature during sample synthesis (Fig. S10). In light of this, the enhanced FM moment, experimentally observed with increasing S deficiency in the bulk [23], is most likely a consequence of the nucleation of the magnetic polarons at the S vacancies.. Moreover, manipulating the SOP density and the total magnetic moment could tune the momentum separation of the Weyl points and the stability of the Weyl semimetal. A higher SOP density ramps up the scattering of the carriers by the spin-orbit exchange field as well as the exchange interaction between the SOPs, unleashing the potential for more robust time-reversal-symmetry-breaking topological phenomena such as the anomalous Hall and Nernst transport at higher temperatures. Given the role of magnetic dopants in dilute magnetic semiconductors [24–27], the vacancy-induced SOP may provide a new path toward generating large magnetic moments in correlated nonmagnetic topological semimetals. In analogy to the NV centers in diamond, SOPs could also provide magnetic moments encoded with topological features as the building blocks for future quantum devices.

Figure 1

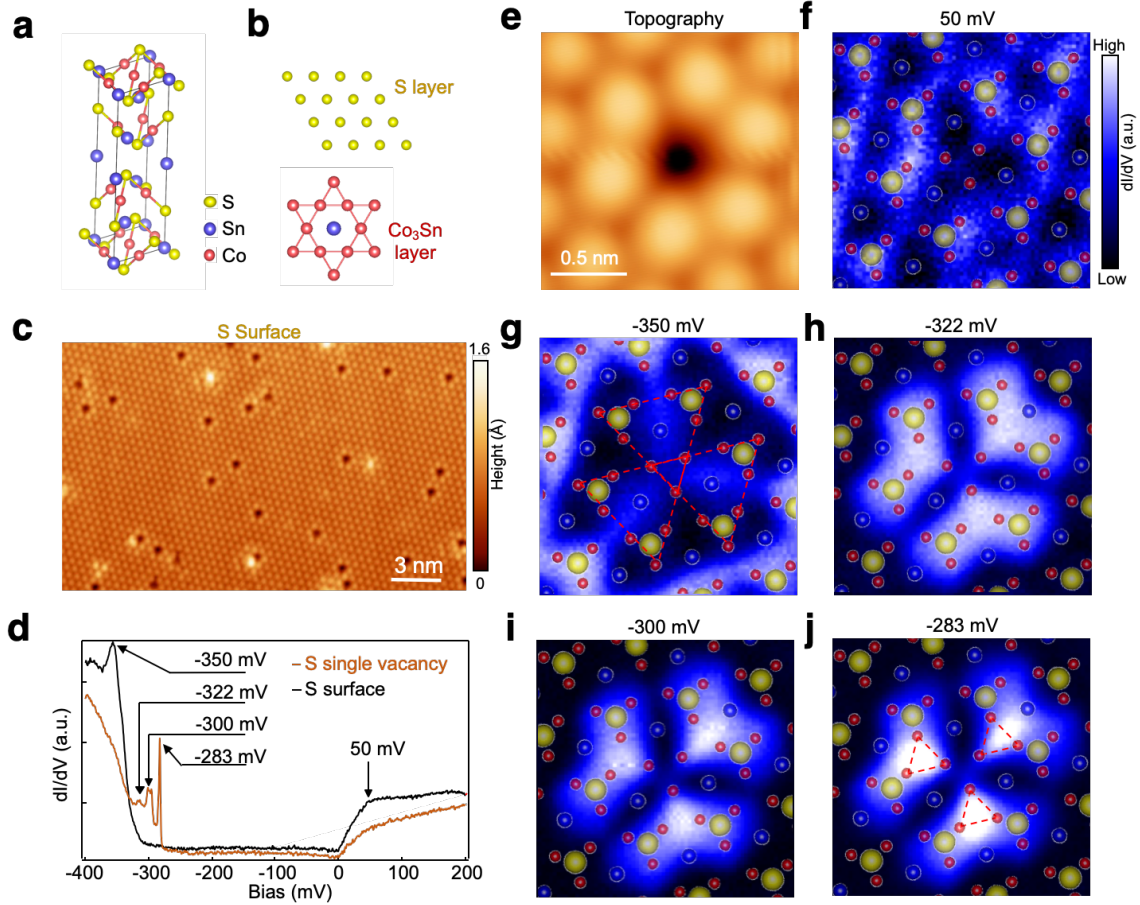

Figure. 1. Localized excitations around a single S vacancy at the S-terminated surface of $Co_3Sn_2S_2$. **a**, Atomic structure of $Co_3Sn_2S_2$. **b**, Atomic structures of S-terminated layer and $Co_3Sn$ kagome layer. **c**, Atomic-resolution STM image of the S-terminated surface, showing randomly distributed single vacancies (scanning setting: bias: $V_s$=-400 mV, setpoint $I_t$=100 pA). **d**, dI/dV spectra at (orange curve) and off (black curve) a single S vacancy ($V_s$=-400 mV, $I_t$=500 pA, $V_{mod}$=0.5 mV). **e**, An STM image of a S vacancy ($V_s$=-400 mV, $I_t$=500 pA). **f-j**, dI/dV maps of (**e**) at different energies: 50 meV (**f**), -350 meV (**g**), -322 meV (**h**), -300 meV (**i**), and -283 meV (**j**), respectively. Atomic structure of $Co_3Sn_2S_2$ is overlay on each map, showing the correlation between the atomic structure and pattern in dI/dV map ($V_s$=-400 mV, $I_t$=500 pA, $V_{mod}$=0.5 mV).

Figure 2

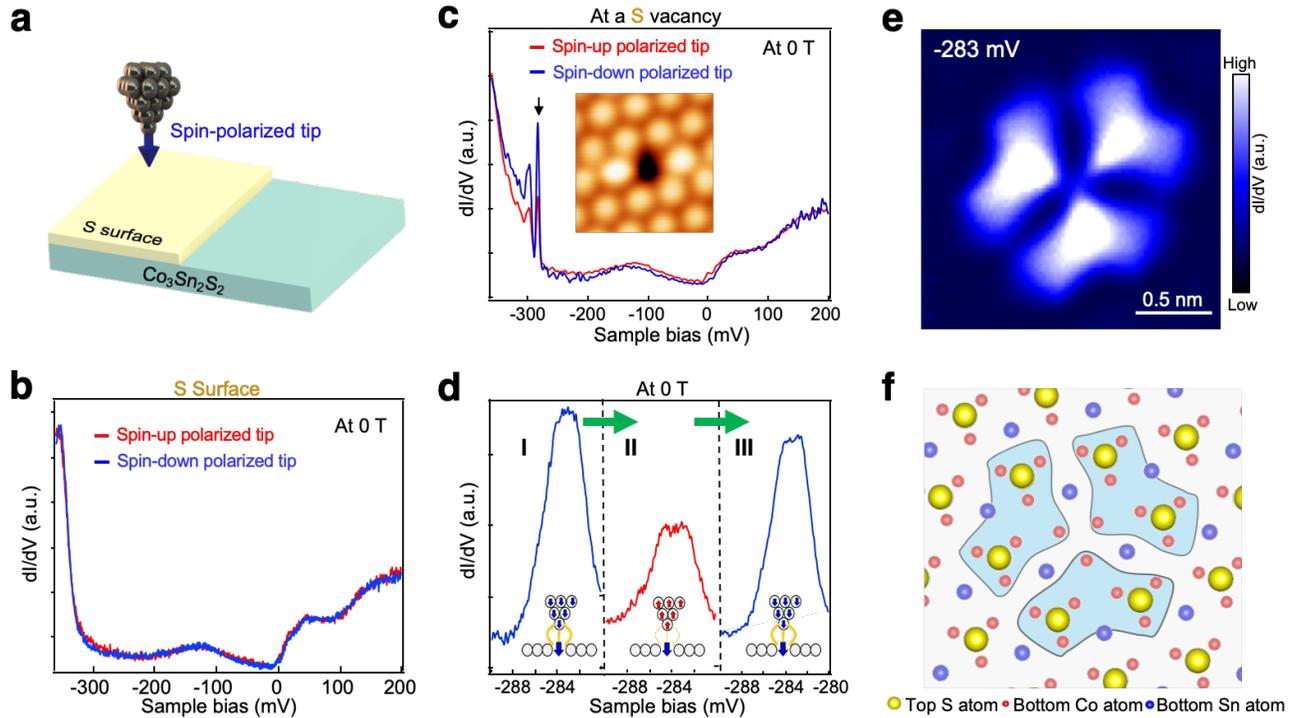

Figure. 2. Spin-polarized bound states at a single S vacancy. **a**, Schematic of spin-polarized measurements using a ferromagnetic Ni tip. **b**, Spin-polarized dI/dV spectra at the vacancy-free region on S-terminated surface using up-polarized tip (red curve) and down-polarized tip (blue curve), showing nearly no polarization contrast on pristine S-termined surface ($V_s$=-400 mV, $I_t$=100 pA, Modulation $V_{mod}$= 0.5 mV). **c**, dI/dV spectra at the center of a single S vacancy, showing a spin-down majority behavior. The inset shows the STM image (2 nm x 2 nm) of the single-S vacancy ($V_s$=-400 mV, $I_t$=200 pA, $V_{mod}$=0.5 mV). **d**, Spin-flip operation of the STM tip and the reproducible spectra at the S vacancy site. Left curve (curve I) corresponds to the initial spin-down tip polarization, the middle one (curve II) corresponds to spin-up tip polarization induced by a magnetic field, and the right one (curve III) corresponds to flipping the spin of the tip back to the initial spin-down polarization. ($V_s$=-400 mV, $I_t$=500 pA, $V_{mod}$=0.5 mV). **e**, dI/dV map of (**a**) at -283 mV, showing that the three-fold rotation symmetry of the bound states localized around the vacancy ($V_s$=-400 mV, $I_t$=500 pA, $V_{mod}$=0.5 mV). **f**, Correlation between the atomic structure and the pattern in the dI/dV map in (E), showing that the spatial distribution of bound states is correlated to the underlying Co atoms.

Figure 3

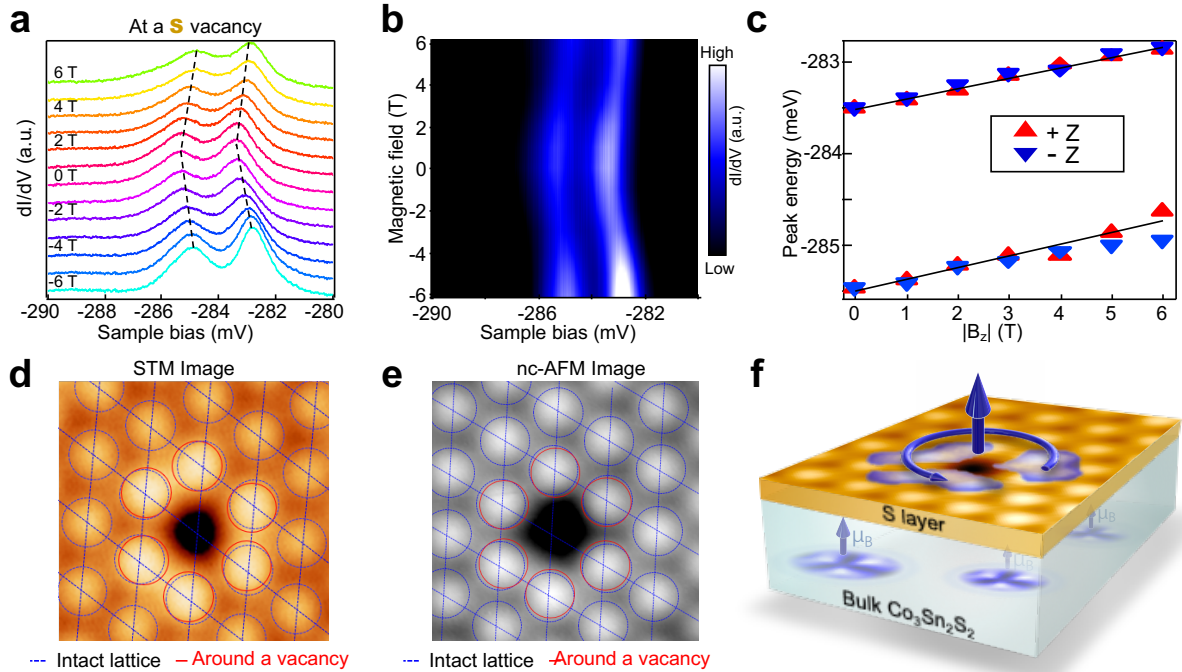

Figure. 3. Anomalous Zeeman shift of the bound states and spin-orbit polaron at a single S vacancy. **a**, dI/dV spectra of the bound states in a magnetic field perpendicular to the sample surface from -6 T to 6 T, showing an approximately linear shift independent of the magnetic field direction. ($V_s$=-400 mV, $I_t$=500 pA, $V_{mod}$=0.5 mV). **b**, Intensity plot of (**a**). **c**, Energy shift of peak positions plotted as a function of the absolute value of magnetic field. **d-e**, STM and nc-AFM images (2.3 nm × 2.3 nm) of the same single-S vacancy, showing appreciable lattice distortion around the single S vacancy (STM: $V_s$=-400 mV, $I_t$=500 pA; AFM: constant frequency shift = -6 Hz). (**f**) Schematic illustration of the localized spin-orbit polaron in $Co_3Sn_2S_2$.